\documentclass[pre,showpacs,twocolumn]{revtex4}
\usepackage{amssymb,amsmath,graphicx}
\usepackage{threeparttable}

\begin{document}

\title{Intelligent tit-for-tat in the iterated prisoner's dilemma game}
\author{Seung Ki Baek}
\author{Beom Jun Kim}
\email[Corresponding author; ]{beomjun@skku.edu}
\affiliation{Department of Physics, BK21 Physics Research Division,
and Institute of Basic Science, Sungkyunkwan University, Suwon 440-746, Korea}

\begin{abstract}
We seek  a  route to the equilibrium where all the agents cooperate in the
iterated prisoner's dilemma game on a two-dimensional plane, focusing on
the role of tit-for-tat strategy.
When a time horizon, within which a strategy can recall the
past, is one time step, an equilibrium can be achieved as cooperating
strategies dominate the whole population via proliferation of tit-for-tat.
Extending the time horizon, we filter out poor strategies by 
simplified replicator dynamics and observe a similar evolutionary pattern
to reach the cooperating equilibrium. In particular, the rise of a modified
tit-for-tat strategy plays a central role,  which implies how a robust
strategy is adopted when provided with an enhanced memory capacity.
\end{abstract}

\pacs{02.50.Le, 87.23.Kg, 89.75.-k, 87.23.Ge}


\maketitle

\section{Introduction}

One of the main interests in statistical physics is related with
the equilibration process of a given system composed of
many interacting elements.
For instance, the classical Ising system made up of locally interacting
spins approaches an equilibrium, characterized by the minimum
of the Helmholtz free energy. Such a model system in statistical
physics is defined by the Hamiltonian and can be readily 
studied by updating spins with local Monte Carlo rules in numerical
simulations~\cite{Newman}.
A lot of interactions including ecological, social, and economical
ones are more complicated than that of spins as they are usually
asymmetric and history dependent. Furthermore, most of these systems
beyond the simple physics model cannot be described by the simple
Hamiltonian approach.
Even if no analytical solution is
available, we may expect the system to evolve by successive local
adaptations, with searching for an optimal point on the fitness landscape.
However, when the interaction is asymmetric, it is possible that
the equilibrium reached by local dynamics may not be optimal in a global
sense.

The prisoner's dilemma (PD) game is a famous model of such disparity.
The typical story begins as follows. Two suspected accomplices are
caught by the police for a crime deserving of 4 years' imprisonment each.
After separating two suspects from each other, the police offers
a deal to each of them:
If only one confesses the crime and the other remains silent, 
the informer will be rewarded and set free, while the other one 
will receive an  aggravated punishment (say 5 years in prison). 
On the other hand, if both keep silent, they will get some punishment which
is supposed to be not so heavy (e.g., 2 years in prison). 
It is still true that they can get light punishments by  cooperating
to each other. From an individual viewpoint, however, it is always better
to defect the other, so they will be eventually sentenced 8 years in
total as the police wanted.
Throughout the present paper, 
the game results are quantified by four elementary payoffs:
The temptation to defect as $T=5$, the reward of cooperation as $R=3$, the
punishment from mutual defection as $P=1$, and the damage from being sucked
as $S=0$. 
Note that the payoffs satisfy two inequalities. The first one $T>R>P>S$
locates the Nash equilibrium~\cite{Nash} at mutual defection, and the second
$2R > T+S$ sets the mutual cooperation as optimal in total.

The conclusion of the PD game is highly nontrivial in that local
optimization will end up with the poorest result in a global sense.
The first breakthrough in this dilemma was made by performing the game
iteratively, where the system could achieve the optimal point of mutual
cooperation~\cite{Axelrod}.
Iteration affects the system's trajectory in two ways:
Since the strategy space comes to have a much larger dimensionality than
choosing between cooperation and defection, there enters a possible route
to mutual cooperation. In addition, as the time scale of interaction
is separated from that of selection, the stability of equilibria and their
basins of attraction may be changed: As to the PD game, for example,
slow selection favors the weaker strategy (i.e., cooperation)
from a population genetics point of view~\cite{Roca}.
Nevertheless, the equilibrium in which all agents cooperate
is usually accessed by a detour consisting of intermediate stages.

In the iterated PD game, there are successfully cooperating strategies some
of which are as follows:
(i)~Grim trigger (GT) initially cooperates, but any single
defection by its opponent makes GT defect forever~\cite{Friedman}.
(ii)~tit-for-tat (TFT) also starts with cooperation, and then
does what the opponent did. This simple strategy is famous for its own
virtues, i.e., being nice, retaliating, forgiving, and
nonenvious~\cite{Axelrod}.
By nice, we mean that a strategy never provokes the opponent first
by defection. Likewise, retaliating and forgiving mean that it defects
after defected, and cooperates when the opponent changes back to
cooperation. Finally, by being nonenvious, TFT allows coexistence of
other strategies. However, one should note that an erroneous defection
between TFT's leads to a chain retribution until a new error makes them
cooperate again.
(iii)~Pavlov keeps its last move if paid highly
and switches to a different move otherwise,
as it is often called win-stay lose-shift~\cite{Kraines}.
Unlike the other two, it forgives a mistake between themselves.

Since the above-mentioned three strategies remember only the moves in
the last time step, they all belong to a set of strategies which
are confined in the time horizon of one time step, which we will call
$M_1$.
Note that the actual amount of information in use is different: GT and TFT
require only the opponent's last move, while Pavlov recalls both of its
opponent's and its own.
Likewise, $M_n$ means the set of strategies which uses the last $k
(\le n)$ time steps in making a decision.
By giving an explicit restriction to the time horizon, our strategy space
is different from that in the state space approach~\cite{Leimar}.

In order to investigate how the system is evolved by selection and
adaptation, we start with every possible strategy in $M_1$ and $M_2$,
respectively, and examine surviving strategies to understand the route to
the equilibrium.
While the genetic algorithm has been often used in exploring a large
strategy space \cite{Axelrod1987,Miller}, we aim at an almost exhaustive
search in that all the strategies are explicitly considered at
least once. In particular, we do not include any mutation processes as
in Ref.~\cite{Lindgren1994} for fixing the strategy space we must
scan. Nor do we treat stochastic
strategies~\cite{Nowak1990,Nowak1992a,Nowak1993,Hauert},
as the deterministic representation shows the pure decision
characteristics of a strategy more clearly. 
Note that we mostly employ typical setups
except for the time horizon in order to keep the situation as simple as possible.
We therefore pass over many interesting variations of the PD
game, such as the idea of payoff-based strategies~\cite{killing}.
The spatial structure we study here is a two-dimensional plane which provides
spatial reciprocity for cooperators~\cite{spatial} (see,
e.g., Refs.~\cite{network,szabo} for other topological
structures), but we do not employ the dynamic preferential
selection~\cite{sheng} and let each agent play with its every
neighbor equally.
Under such conditions, we find that $M_2$ has its own TFT modified from the
original one in $M_1$, which seemingly indicates a generic pattern in the
evolution of cooperation.
Even though the reciprocity has been thought of as relevant to the
emergence of cooperation even in longer time horizons~\cite{Axelrod1987},
such concrete strategic forms, which are directly related to the
original TFT, have not been reported yet.

The present paper is organized as follows:
In Sec.~\ref{sec:methods}, we check the case of $M_1$ to introduce our basic
scheme. In Sec.~\ref{sec:appl}, we apply it to $M_2$ and present the
surviving strategies, including the modified type of TFT. Finally, we
discuss and conclude this work in Sec.~\ref{sec:summary}. 

\section{Methods}
\label{sec:methods}
\subsection{Bitwise representation of strategies in $M_1$}
A strategy in $M_1$ can be conveniently denoted by five bits,
each of which can take either cooperation ($C$) or
defection ($D$):
The first bit, $\alpha$, is the move when a player
first encounters an opponent and thus has empty memory.
The bit $a_1$ is the move at time $t$ when the player's
and opponent's previous moves at $t-1$  were $C$ and $C$
[henceforth we denote this situation as
(player's move at $t-1$, opponent's move at $t-1$) = $(C,C)$] ,
respectively (Table~\ref{table:str}).
Likewise, $a_2$ is for $(C,D)$, $a_3$ for $(D,C)$, and
$a_4$ for $(D,D)$.

\begin{table}
\caption{Bitwise representation of a strategy in $M_1$ as
$\alpha|a_1 a_2 a_3 a_4$.}
\begin{tabular*}{\hsize}{@{\extracolsep{\fill}}lcccccc}
\hline\hline
State & Empty & ($C$,$C$) & ($C$,$D$) & ($D$,$C$) & ($D$,$D$) &\\
Player's move      & $\alpha$ & $a_1$ & $a_2$ & $a_3$ & $a_4$
&\\\hline\hline
\end{tabular*}
\label{table:str}
\end{table}

Consequently, a strategy in $M_1$ is coded
by $\alpha|a_1 a_2 a_3 a_4$ and the total number of strategies
is $|M_1| = 2^5=32$,
for each of five bits can have either $C$ or $D$.
For example, $C|CDDD$, $C|CDCD$, and $C|CDDC$
encode GT, TFT and Pavlov, respectively.
Further examples include
the unconditional cooperator (ALLC or AC)
coded by $C|CCCC$ and the unconditional
defector (ALLD or AD) by $D|DDDD$.  A nice strategy (see above)
in $M_1$ is represented as $\alpha = C$, implying that
it starts with $C$ at the first encounter, and $a_1 = C$, meaning
that it never provokes the defection first
\footnote{The use of the bit notation turns out to be
very convenient in our actual computer programming, in which we substitute
$C$ to 1 and $D$ to 0.  Thus, for example, TFT is represented as a
binary number $11~010$, corresponding to the decimal integer 28. In this way,
any strategy in $M_1$ is written as an integer $i = 0, 1, \ldots,
31$.}.

\subsection{Transition graphs and tournament for $M_1$}
\begin{figure}
\includegraphics[width=.22\textwidth]{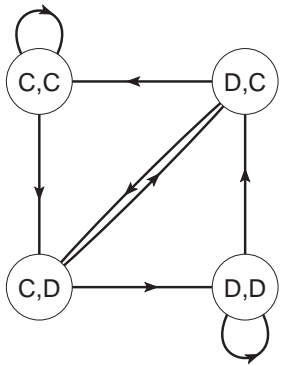}
\caption{ Transition graph for TFT.
Each vertex represents a state in Table~\ref{table:str}, and the directed
edges are the possible  next states allowed by this strategy.
Each vertex has two outgoing arcs, considering the move taken by the
opponent.}
\label{fig:tran}
\end{figure}

Another way of representing a strategy is to mention all of the possible states
it may meet and all of the possible transitions between
them~\cite{Miller,Ashlock,Sethi}.
Identifying each state with a vertex and each
transition with an arc (a directed edge), with self-connecting included,
this procedure yields 
a transition graph for each strategy.
Suppose, for example, that Alice employs TFT and Bob does another arbitrary
strategy in $M_1$. From Alice's viewpoint, the four possible states are
represented by four pairs; $(C,C)$, $(C,D)$, $(D,C)$, and $(D,D)$ where the
former character indicates her last move and the latter does Bob's. If
starting with $(C,C)$, the next state must be $(C,X)$ with $X=C$ or $D$
depending on Bob's strategy, because Alice remembers what Bob did 
at the last encounter. Repeating this for all the states gives 
Fig.~\ref{fig:tran}. One can easily get the graphical representations
for any other strategies by the same procedure, noting that the initial bit
$\alpha$ does not change the transition graph but only makes the starting
vertex in the graph different.

From all the 16 transition graphs in $M_1$,
TFT is found to be unique in that it does not permit returning
to $(C,D)$ without visiting $(D,C)$, which implies that any strategy
cannot repeatedly suck TFT avoiding retaliation.
In order to describe the time course of the PD game between two
agents, the distinction between transient and recurrent
states
needs to be made: Transient states have only outward
arcs and thus cannot be visited repeatedly, while the
recurrent states are visited over and over again. For example, TFT
does not have transient states, while AD has two transient states
$(C,C)$ and $(C,D)$ with two recurrent states $(D,C)$ and $(D,D)$.

\begin{figure}
\includegraphics[width=.22\textwidth]{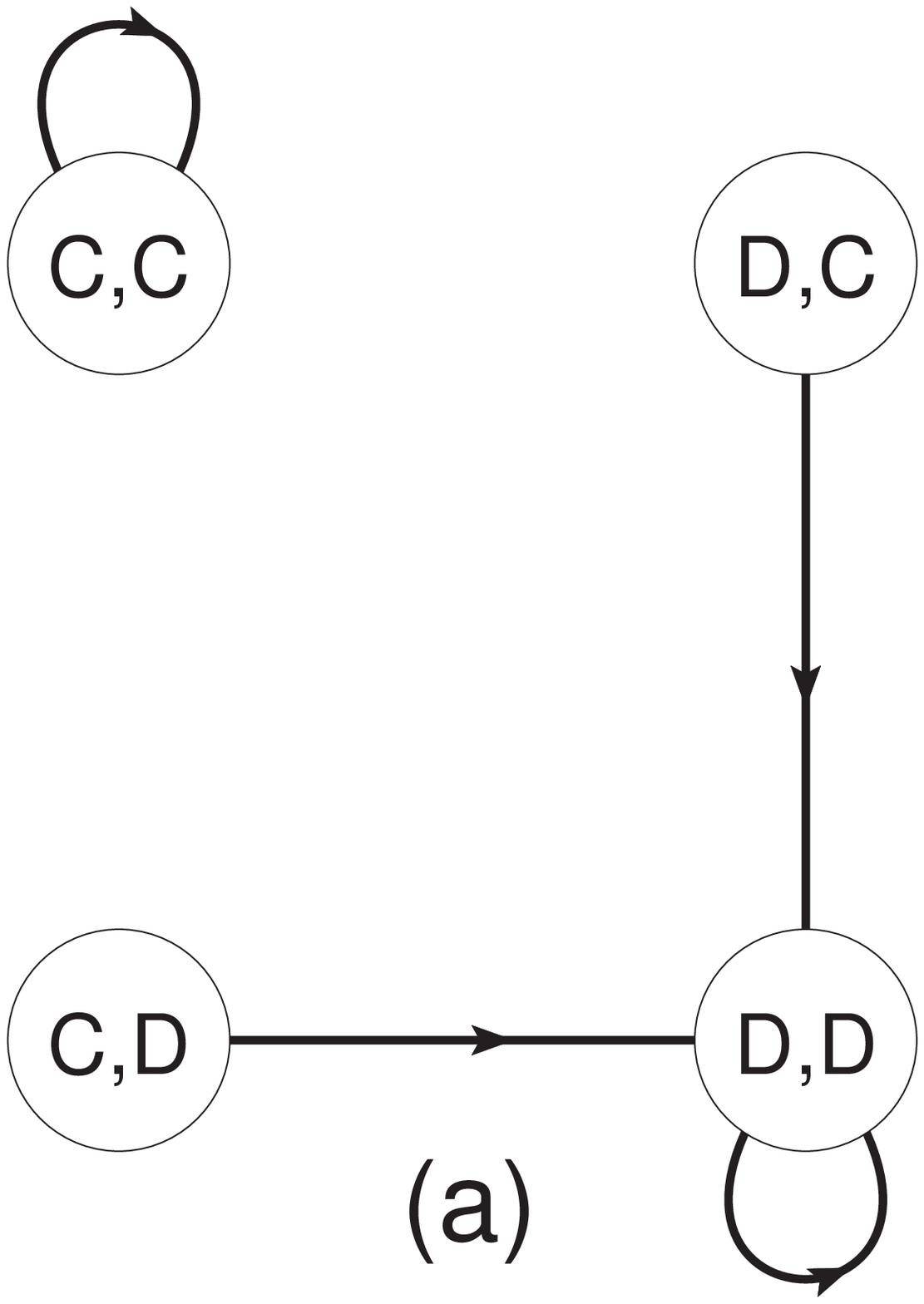}
\includegraphics[width=.22\textwidth]{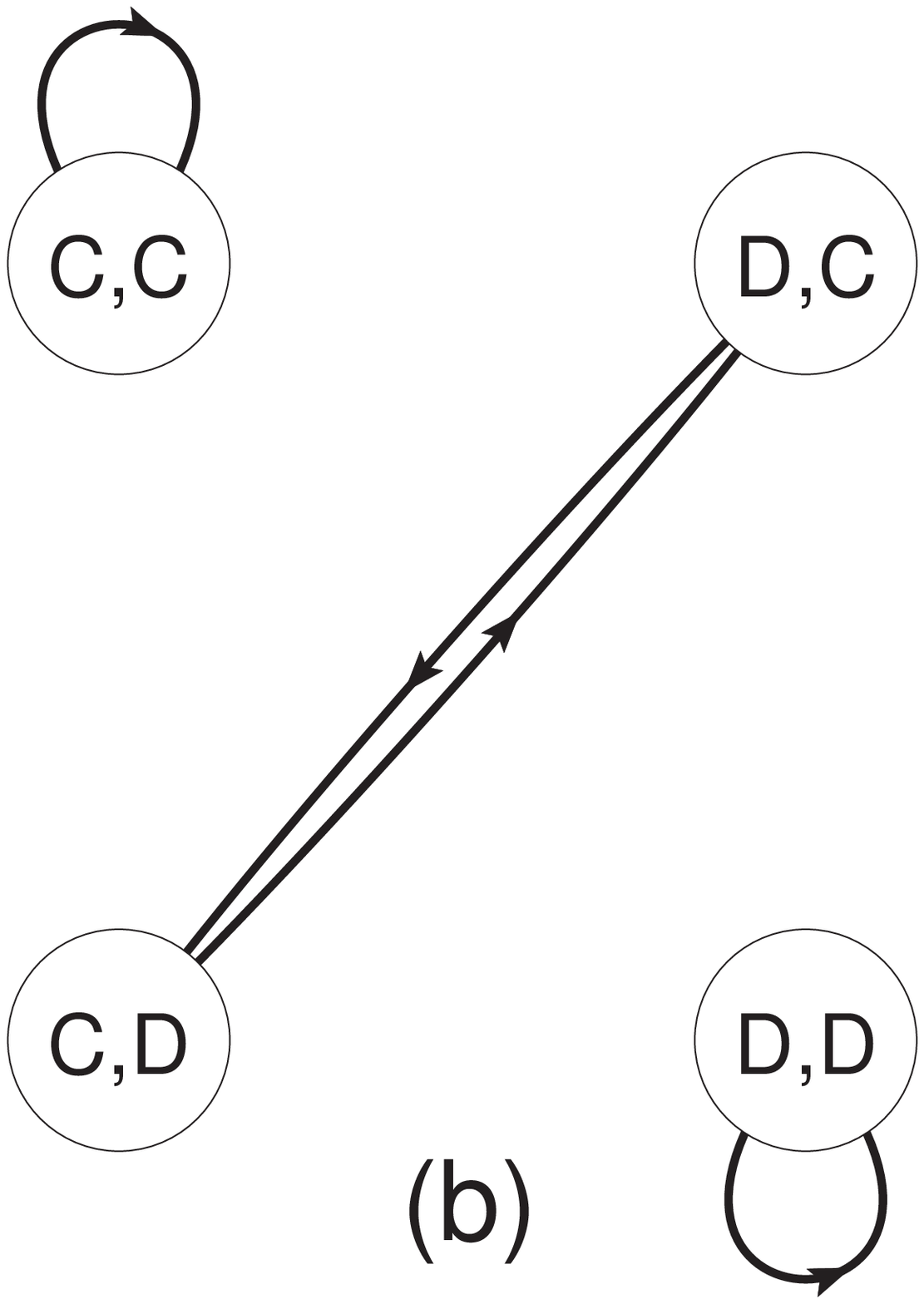}\\
\includegraphics[width=.22\textwidth]{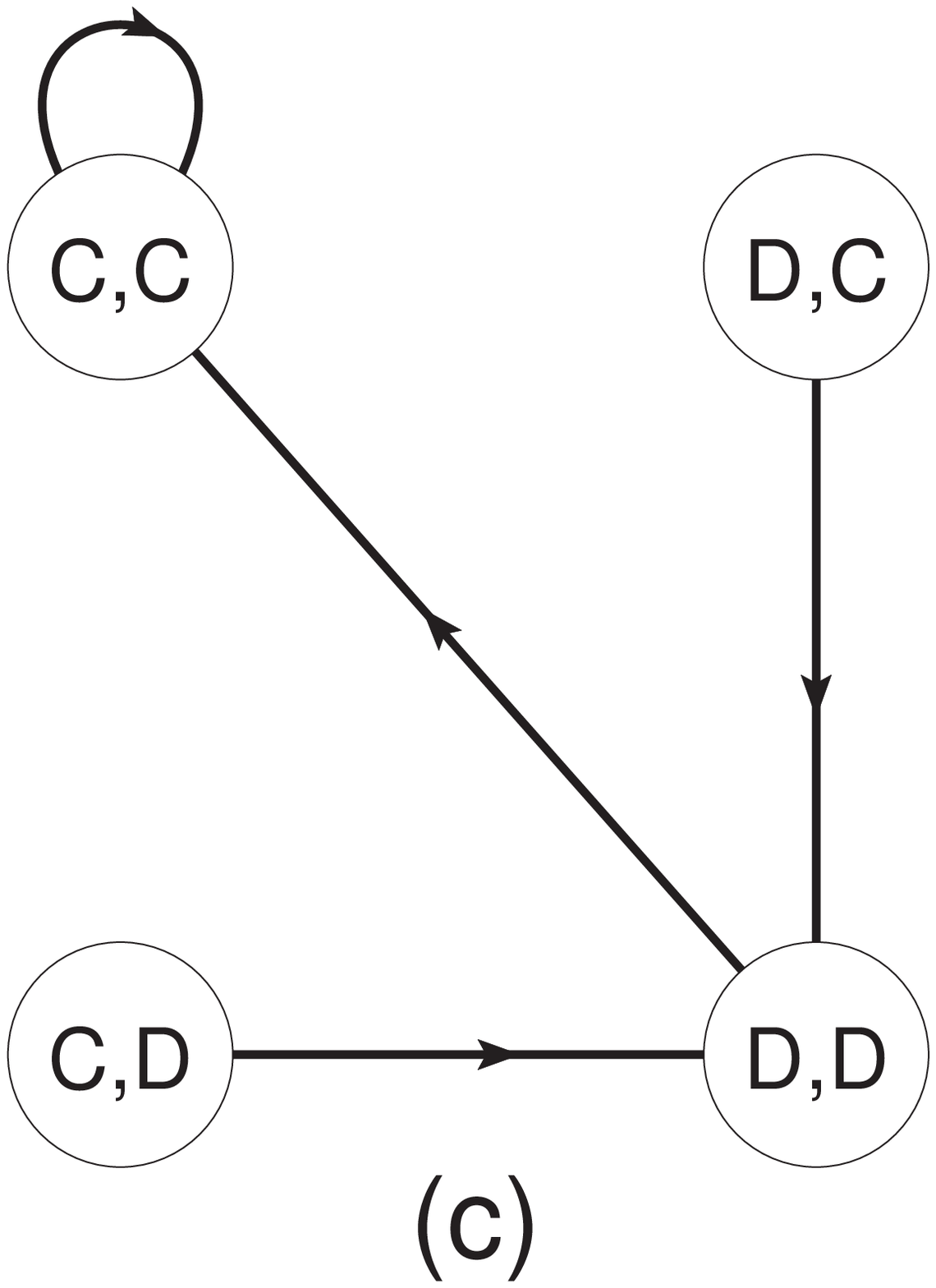}
\includegraphics[width=.22\textwidth]{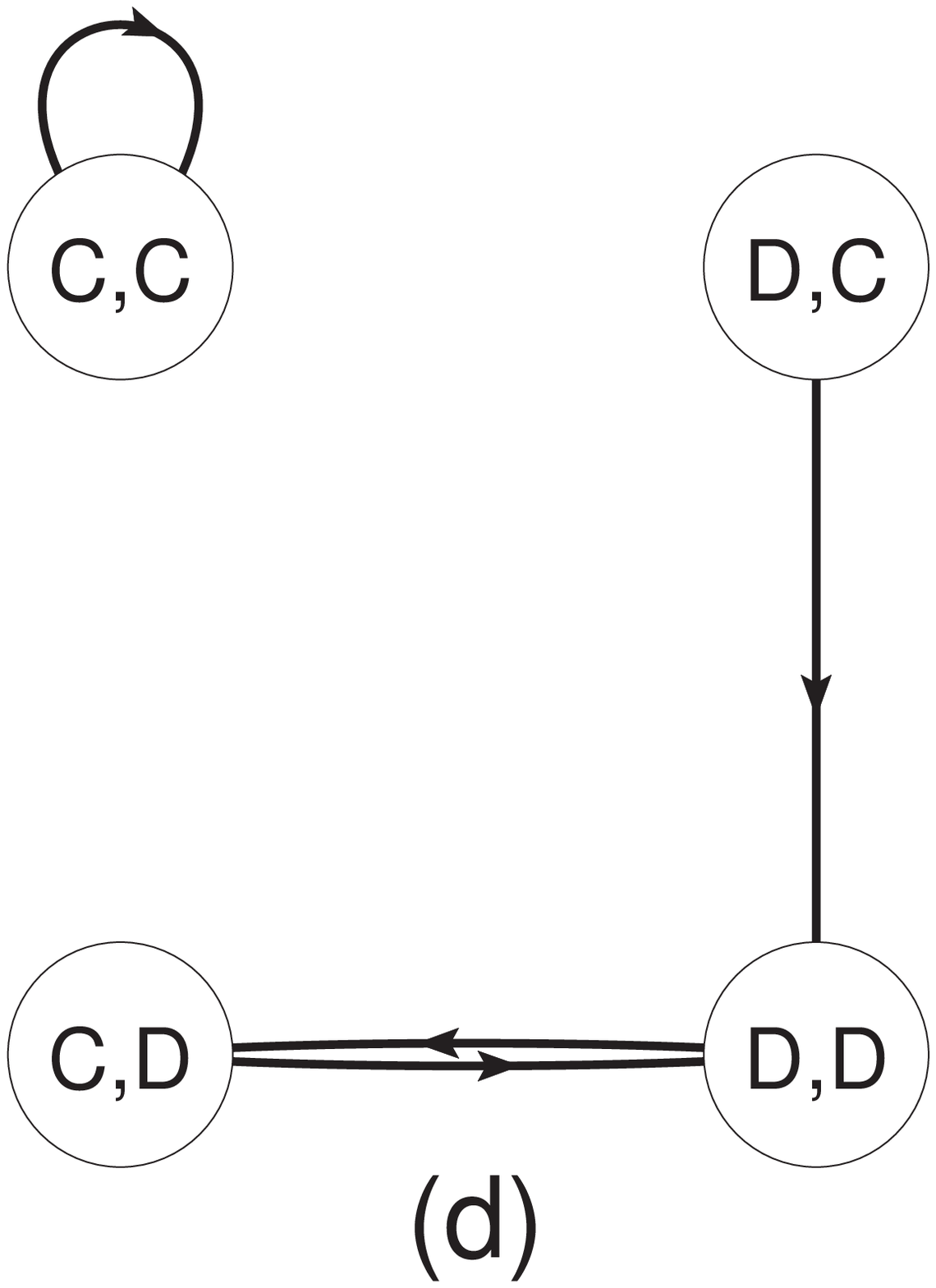}
\caption{
Transition graphs from combining two strategies in $M_1$.
(a) GT vs GT. If deviated from $(C,C)$, the only attractor is mutual
defection. (b) TFT vs TFT. If mistaken, they do not recover mutual
cooperation on their own, unless another error brings them back.
(c) Pavlov vs Pavlov, forgiving an error
between themselves. (d) Pavlov vs GT. Pavlov is defeated by GT if any
error occurs.}
\label{fig:match}
\end{figure}

If two strategies $i$ and $j$ in $M_1$ play the PD game
together, two corresponding graphs are combined to make one deterministic
transition graph (Fig.~\ref{fig:match}).
The move sequence is periodic
and the long-time limit of the average payoff per time step
is determined only by recurrent states of the two,
from which one can calculate
easily $U_{ij}$, the average payoff per step that the strategy $i$ gains
from $j$.
In the same spirit as the original tournament held by 
Axelrod, we compute the average points
the strategy $i$ gets from all strategies
(including $i$) to obtain Table~\ref{tab:M1}.
So far as each pair of strategies has an equal acquaintance probability,
the tournament results will converge to these values in the long-time
limit.
Moreover, since each value in this table is analytically calculated from
periodic moves in pairs of deterministic strategies, one can decompose it
into the elementary payoffs, $T$, $R$, $P$, and $S$. For example, AD, AC,
and GT earn $(T+P)/2 = 3$, $(R+S)/2 = 1.5$, and $R/2 + (T+P)/4 = 3$,
respectively.
One can see that
TFT is not the best within $M_1$ and that strategies with more $D$ bits
often outperform cooperators. We emphasize that
the above results in a round-robin tournament are not related to an
evolutionary process yet and need to be checked from evolutionary
perspectives.

\begin{table*}
\caption{Average points $\frac{1}{|M_1|}\sum_j U_{ij}$ for each
strategy in $M_1$.}
\begin{tabular*}{\hsize}{@{\extracolsep{\fill}}lc@{\hspace{1cm}}lc@{\hspace{1cm}}lc@{\hspace{1cm}}lc}
\hline\hline
Strategy & Points  & Strategy & Points  & Strategy & Points  & Strategy &
Points \\
\hline
AD       & 3.00 & $C|DDDC$ & 2.73 & $C|DDCC$ & 2.25 &  $C|DCCD$ & 1.69 \\
$D|CCDD$ & 3.00 & $D|DCDC$ & 2.73 & $D|DDCD$ & 2.22 &  $C|DCCC$ & 1.63 \\
$C|DDDD$ & 3.00 & Pavlov   & 2.56 & $C|CDCC$ & 2.19 &  AC       & 1.50 \\
GT       & 3.00 & $D|CCDC$ & 2.38 & $D|CDCC$ & 2.09 &  $C|CCDC$ & 1.50 \\
$D|CDDD$ & 3.00 & $C|DDCD$ & 2.36 & $D|CDCD$ & 2.02 &  $C|CCCD$ & 1.50 \\
$D|DCDD$ & 3.00 & TFT      & 2.35 & $C|DCDC$ & 1.90 &  $C|CCDD$ & 1.50 \\
$D|DDDC$ & 2.97 & $D|DDCC$ & 2.25 & $D|DCCD$ & 1.86 &  $D|CCCC$ & 1.50 \\
$D|CDDC$ & 2.89 & $C|DCDD$ & 2.25 & $D|DCCC$ & 1.81 &  $D|CCCD$ & 1.38 \\
\hline\hline
\end{tabular*}
\label{tab:M1}
\end{table*}

\subsection{Spatial prisoner's dilemma game for $M_1$}
The spatial PD game (SPDG) provides a good framework
for observing the emergent cooperation as it allows the cooperating
strategies to make clusters against
defectors~\cite{spatial}.
There is no unique standard in constructing SPDG, and a different rule
may yield a different output, in general. Here we present our SPDG rules,
which have been extensively used in literature~\cite{szabo}.

We perform SPDG on a
two-dimensional $128 \times 128$ square lattice with the periodic
boundary condition. In the initial stage of the SPDG, one among
all 32 strategies in $M_1$ is randomly assigned to each node of the
lattice,
and every agent plays the PD game with her four nearest neighbors.
After all agents play the game, often called
one Monte Carlo (MC) step, this procedure is
stopped with a preassigned probability $p$ or repeats itself with $1-p$.
When stopped, the sequence of games so far is termed as
one generation whose average time duration is $1/p$ MC steps.
In order to make the effects of transient states (see above) as weak as
possible, $p$ should be sufficiently small to ensure that one
generation is long enough (we observe that $p=0.05$, corresponding
to one generation as 20 MC steps on average,  fulfills this
requirement).
Whenever a generation is closed, the selection mechanism is activated as
follows:
Every node, one by one, randomly chooses one of its nearest neighbors and
adopts the neighbor's strategy if the neighbor has gained more during that
generation. Memory tables for all pairs of agents are recalculated
and payoffs are initialized back to zero, and then the next generation begins.

\begin{figure}
\includegraphics[width=.48\textwidth]{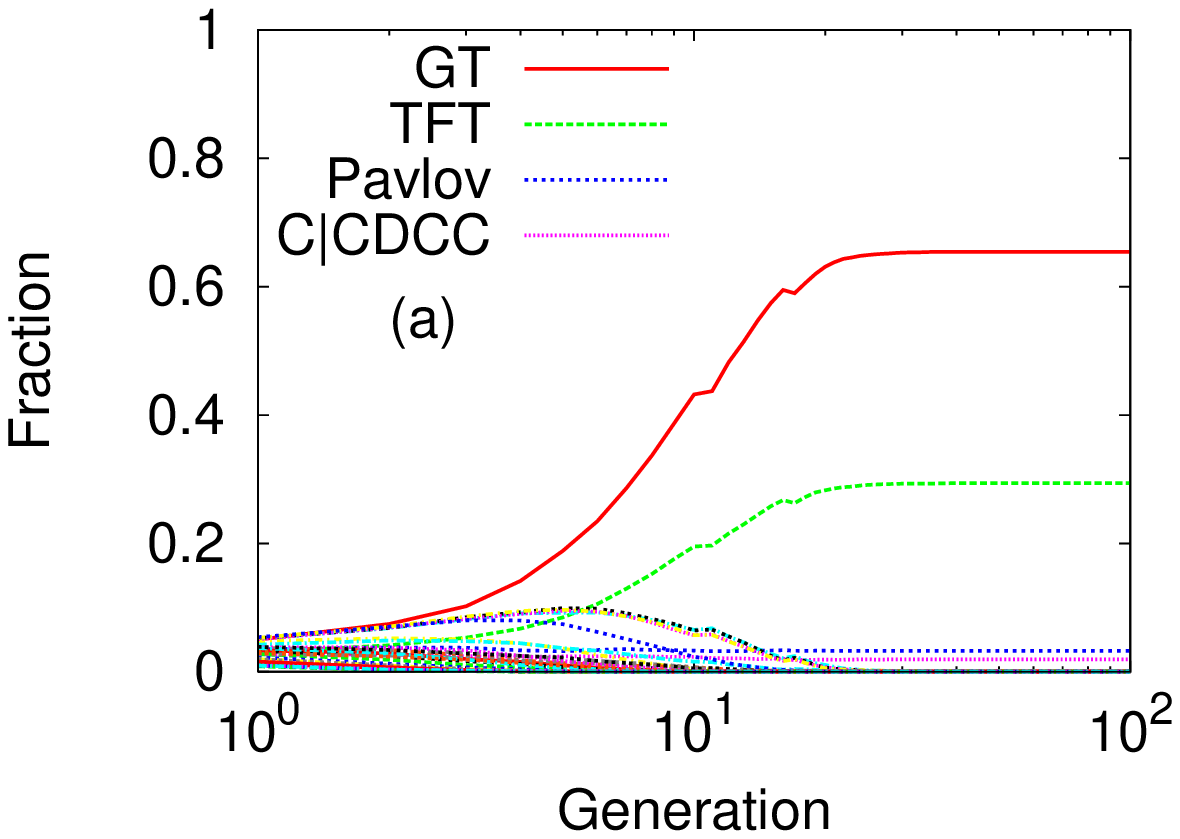}
\includegraphics[width=.48\textwidth]{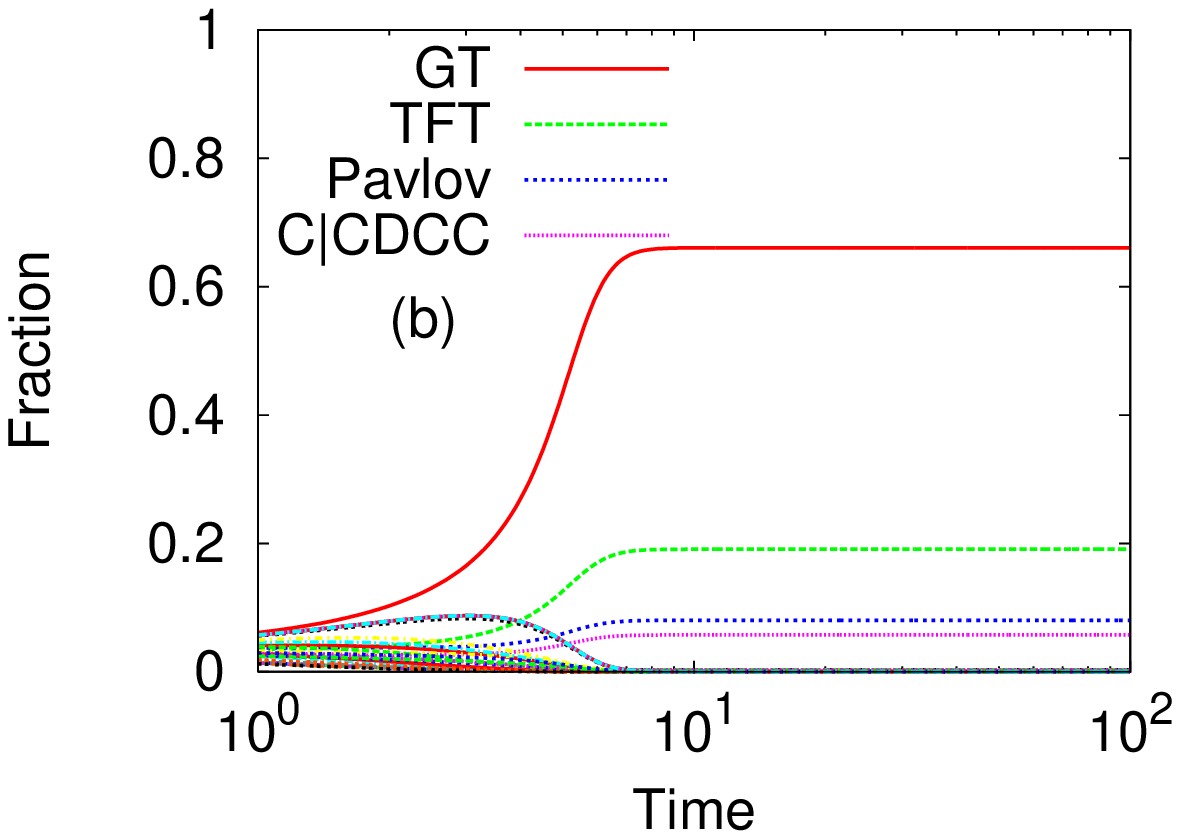}
\caption{(Color online)
Comparison between SPDG and RD.
(a) A simulation result on a $128\times128$ lattice with $p=0.05$.
(b) Numerical integration of Eq.~(\ref{eq:rd}).
}   
\label{fig:ck_fig} 
\end{figure}

\begin{figure}
\includegraphics[width=.48\textwidth]{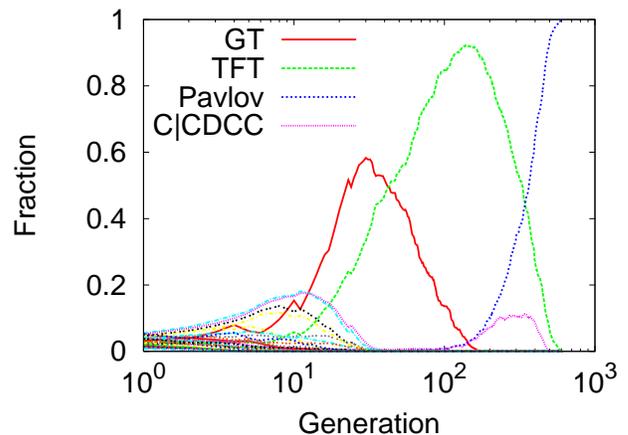}
\caption{(Color online)
A pattern in SPDG on the $128\times128$ lattice
with $e=0.01$ and $p=0.05$. Even with the presence of error, almost
all the dynamical patterns at large time scales occur within the
strategies found in the error-free RD, if the error probability is
sufficiently low.}
\label{fig:ck_cluster}
\end{figure}

Our SPDG simulation readily shows that a cooperating
equilibrium, in which all of the agents are playing $C$,
is achieved mostly by GT, TFT, and Pavlov, together with 
a minor strategy $C|CDCC$ [Fig.~\ref{fig:ck_fig}(a)].
It is notable that these surviving four strategies are,
in fact, the four possibilities when we fix the nice
bits ($\alpha = C, a_1 = C$) and the retaliating bit ($a_2 = D)$.
This implies that the virtues of TFT
(see above) are indeed very important conditions for a strategy to be 
evolutionarily successful.

\subsection{Replicator dynamics and filtering}
As the direct SPDG often requires an amount of computation, we bypass the
problem using the replicator
dynamics (RD)~\cite{Weibull} with average
payoffs~\cite{Kandori,Lindgren1992,Tanimoto}:
Once the average payoffs are obtained from the transition graphs,
the time evolution of the fraction of each strategy can,
phenomenologically but conveniently,
be described by RD within
the assumption of the full mixing, corresponding to the
mean-field approximation.

Suppose that we perform SPDG with randomly distributed strategies
in a two-dimensional $L \times L$ square lattice with the total
number of agents $N_a \equiv L^2$. 
Since each agent plays the game with $z=4$ nearest neighbors, the expected
gain that an agent with the strategy $i$ collects, 
within the assumption of a full mixing, is written as
\begin{equation}
U_i = \sum_j z U_{ij} \phi_j, 
\label{eq:mf_payoff}
\end{equation}
where $\phi_j$ is the fraction  defined as the
number of agents of the strategy $j$ divided by $N_a$ with $\sum_{i} \phi_i
= 1$, and $U_{ij}$ is the above-mentioned average gain $i$ gets from $j$.
If the relative growth rate of a strategy is proportional to its relative
payoff deviated from the average over the whole population, we may write an
ordinary differential equation
\begin{equation}
\frac{d\phi_i}{dt} = \left( U_i - \sum_{j} U_j \phi_j \right)\phi_i,
\label{eq:rd}
\end{equation}
which is called the replicator dynamics.
Note that if each strategy forms a cluster, the summation over the nearest
neighbors of $i$ cannot cover the whole space, and we must examine what
happens near the interfaces.

Although the RD description 
is more crude than the actual SPDG with
local interactions, we find that the
numerical integration of RD is surprisingly 
similar to what SPDG yields with the
random initial distribution of strategies.
In Fig.~\ref{fig:ck_fig}(b), it is displayed that 
the four nice strategies of GT, TFT, Pavlov, and $C|CDCC$
survive just like the previous observation
made for SPDG. Furthermore, the order of relative fractions
of the four is identical in both results.
Note that these four strategies are indistinguishable at this stage,
because the bits other than $a_1$ are not actually used any more.
In order to slightly activate those bits
and check how the surviving strategies behave
in the presence of erroneous decisions, we allow each
player in SPDG to make mistakes at a given probability $e$.
For example, $e=0.01$ means that an agent's memory on a neighbor's
last move may be flipped from $C$ to $D$ or $D$ to $C$, 
once in 100 moves on average.
Depending on the initial condition, various
steady-state configurations are obtained.
In many cases, however, we find that Pavlov eventually
conquers the whole territory, defeating TFT~\cite{Nowak1993},
under such a low error rate (Fig.~\ref{fig:ck_cluster}).

It is important that the error-free RD equilibrium selects out
the long run strategies which appear in SPDG~\cite{Tanimoto}.
The dynamics among these strategies are driven by
errors in much larger time scales than the fast extinctions.
When $e \ll 1$, the difference in these two time scales makes it possible to
separate the fast extinctions from the long run behaviors.
We point out that this selection can be further simplified,
considering that each strategy occupies only a small fraction at the
early stages and that the strategy with the least payoff decreases
most rapidly.
That is, the least fit strategy will be shortly removed from the
population in effect, and the remainder's payoffs are rectified
accordingly.
Eliminating the least fit actually reaches the same cooperating equilibrium
with the minimal number of computations, and
it works similarly to the technique called the iterated
elimination of dominated strategies~\cite{Borg}.
This procedure will be denoted as RD filtering since it is based on a
fundamental assumption of RD that the growth rate of a species is
proportional to its payoff.
After it simulates the initial short times until reaching an equilibrium,
we come back to SPDG and consider the slow dynamics due to errors among
survivors.
Nevertheless, we stress that this procedure is only a rough approximation
and one should be careful not to expect general coincidence between them.
Based on the numerical support in $M_1$, we are suggesting that
this procedure can be regarded as a criterion that a feasible strategy is
supposed to pass, rather than as a precise equivalent of SPDG.
One obvious drawback is that it precludes much of the possibility of
cyclic behaviors allowed by the continuous RD~\cite{cycle}, as we give the
least fit no chance to return back (via, e.g, mutations) once removed.

\section{Application to $M_2$}
\label{sec:appl}
\subsection{Approach to memory effects}
Let us proceed to the study of the strategies in $M_2$
to examine the effects of memory capacity in evolution.
In order to decide the move at time $t$, an agent 
needs to remember her own moves and the opponent's moves at $t-1$ and $t-2$,
respectively, corresponding to $2^4 = 16$ bits. Until the agent meets
the opponent more than once, the past information is not yet available
and thus the strategy should specify the moves for this case with two more
bits for the initial two encounters.
Accordingly, the number of strategies 
in $M_2$ is counted as $|M_2| = 2^{16+2} = 262~144$.
Based on the previous results for $M_1$, 
we use the same method to filter out unsuccessful strategies
in an early stage, 
and then play SPDG only for surviving strategies.

\begin{table*}
\caption{Strategy table for $M_2$.}
\begin{tabular*}{\hsize}{@{\extracolsep{\fill}}ccccc@{\hspace{1cm}}ccccc}
\hline\hline
State$^{\rm a}$ & $\Omega$~($C$/$D$)$^{\rm b}$ & EC$^{\rm c}$
& ET$^{\rm d}$ & I-TFT$^{\rm e}$
& State & $\Omega$~($C$/$D$) & EC & ET & I-TFT
\\\hline
($CC$,$CC$)&100/0&$C$&$C$&\underline{$C$}&($DC$,$CC$)&50/50&$D$&$D$&\underline{$C$}\\
($CC$,$CD$)&42/58&$C$&$C$&\underline{$D$}&($DC$,$CD$)&45/55& - & - &\underline{$D$}\\
($CC$,$DC$)&52/48&$D$&$D$&$C$            &($DC$,$DC$)&50/50&$C$&$D$&$C$\\
($CC$,$DD$)& 6/94& - & - &$D$            &($DC$,$DD$)&47/53& - & - &$C$\\
($CD$,$CC$)&54/46&$C$&$C$&$D$            &($DD$,$CC$)&52/48& - & - &$C$\\
($CD$,$CD$)&48/52&$C$&$C$& - $^{\rm f}$
                               &($DD$,$CD$)&47/53&-&-&$C$\\
($CD$,$DC$)&56/44&-&-&\underline{$C$}&($DD$,$DC$)&50/50&-&-&\underline{$C$}\\
($CD$,$DD$)&31/69&-&-&\underline{$D$}&($DD$,$DD$)&53/47&-&-&\underline{$D$}\\
\hline\hline
\end{tabular*}
\label{table:str2}
{
\begin{tablenotes}
\item $^{\rm a}$ A state $(X_1 X_2,Y_1 Y_2)$ means that $X_1$ and
	$X_2$ ($Y_1$ and $Y_2$) are player's (the opponent's) moves at two
	subsequent times, respectively.
\item $^{\rm b}$ Percentages of $C$ and $D$ in $\Omega$, the set of the
	remaining strategies after RD filtering.
\item $^{\rm c}$ Efficient cooperator's moves at each given state.
\item $^{\rm d}$ Efficient trigger's moves. Although similar to EC's,
	this does not follow EC at $(DC,DC)$ but defects it.
\item $^{\rm e}$ I-TFT's moves. The moves at the recurrent states are
	underlined.
\item $^{\rm f}$ If both of $C$ and $D$ are observed, the move is written as
	blank.
\end{tablenotes}
}
\end{table*}

\subsection{RD filtering}
In the filtering procedure for $M_2$, we again calculate
$U_{ij}$ in the same way as before and use the mean-field payoff function 
in Eq.~(\ref{eq:mf_payoff}), assuming the full mixing. This reflects the
fact that the initial strategies are randomly distributed and
the number of remaining strategies turns out to be large enough to neglect
clustering effects even at the equilibrium.
The iterated elimination stops when no more strategies can be removed.

\begin{figure}
\includegraphics[width=.48\textwidth]{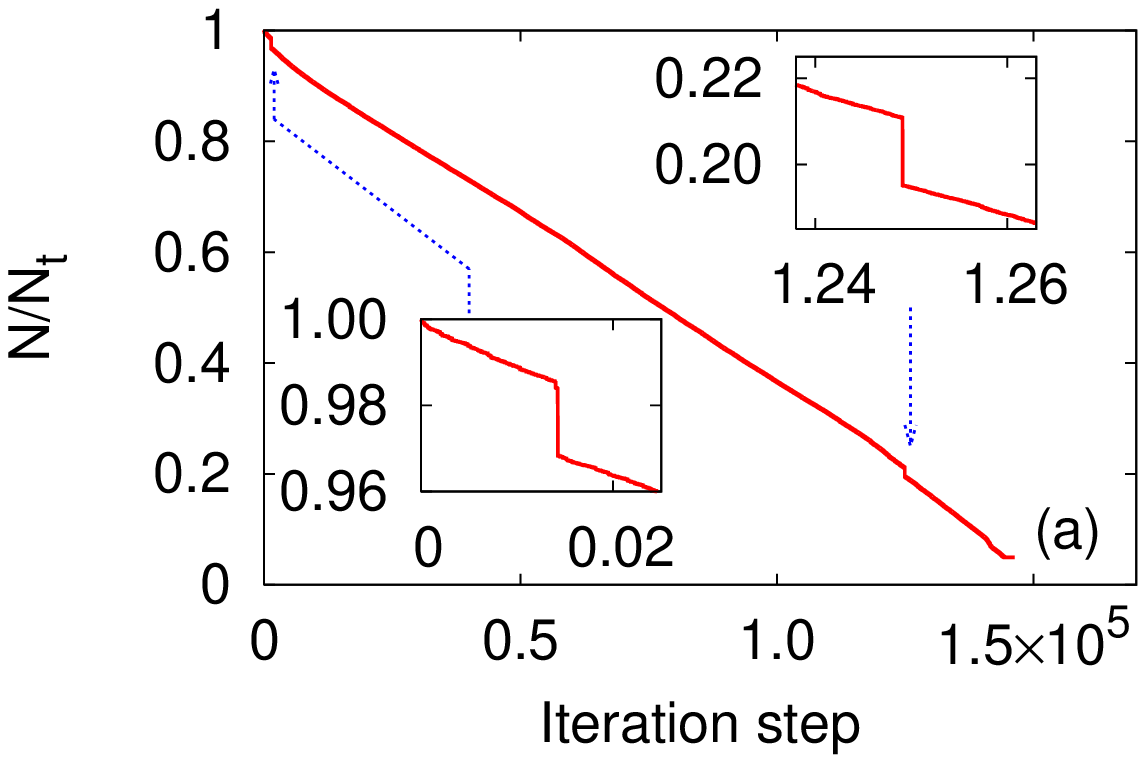}
\includegraphics[width=.48\textwidth]{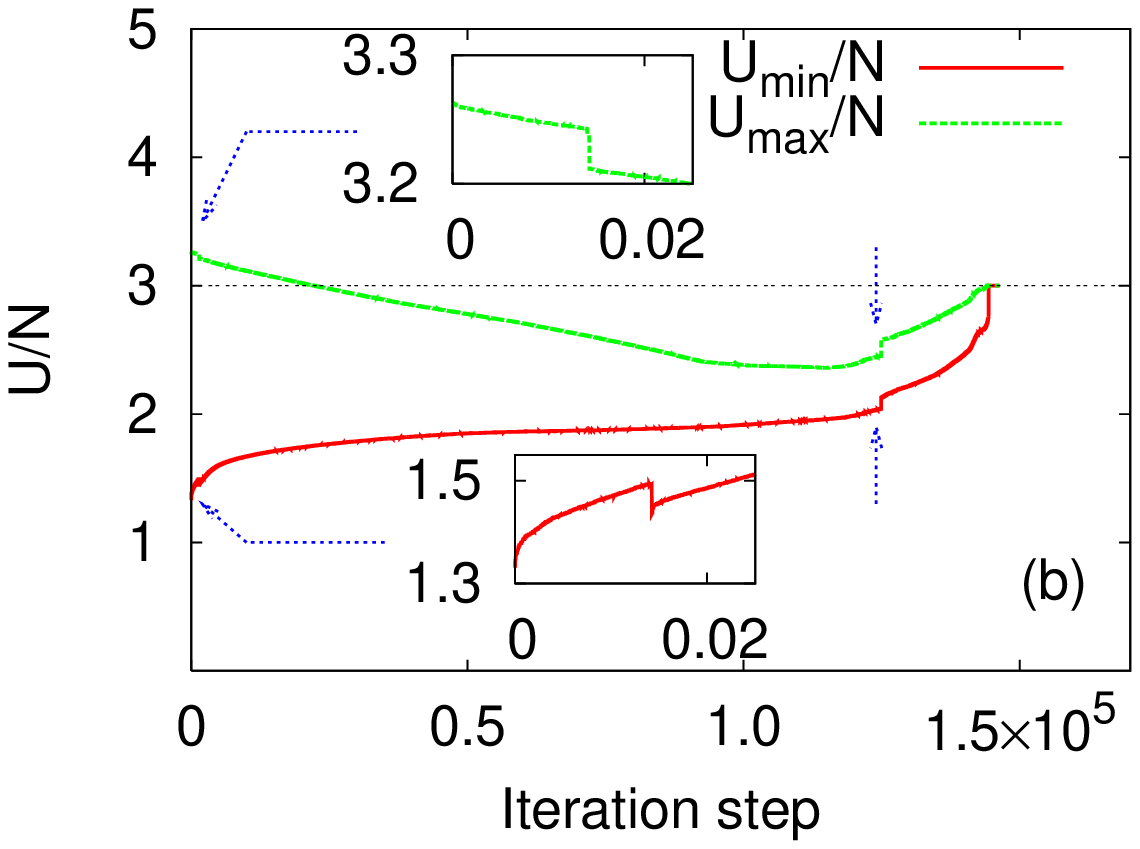}
\caption{(Color online)
Filtering procedure on $M_2$. It takes
about $1.4\times10^5$ steps to reach a cooperating equilibrium.
(a) The remaining fraction $N/N_t$, where $N$ is the number of
survivors and $N_t = 2^{18} = 262~144$. Insets show the two great extinction
events at the $1424$th and $124~910$th steps, respectively.
(b) The maximum and minimum values of the payoffs, divided by $N$.
The straight line represents $R=3$, the reward for mutual cooperation.
Insets and arrows are for the great extinction events again.}
\label{fig:pc}
\end{figure}

\begin{figure*}
\includegraphics[width=0.9\textwidth]{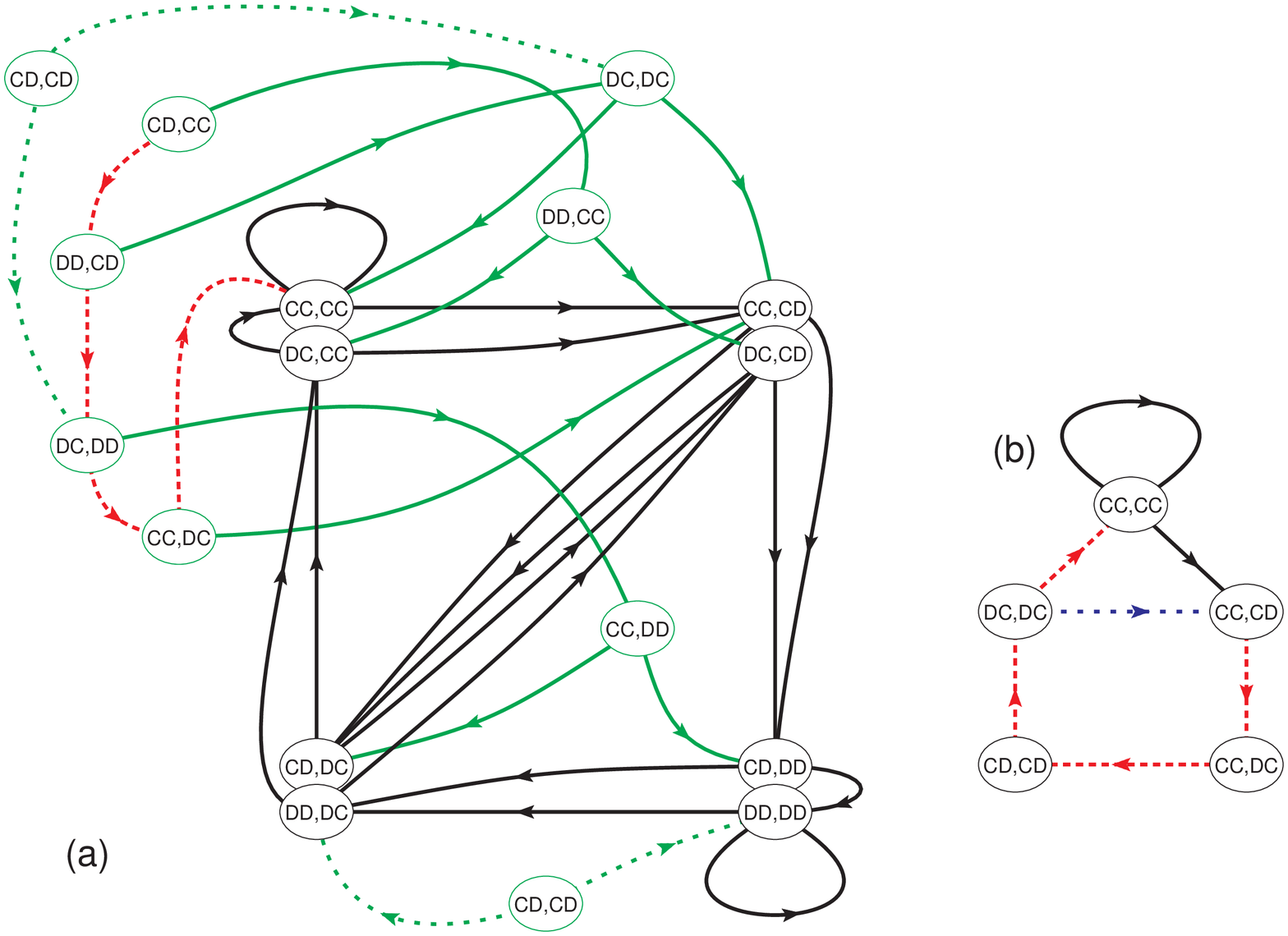}
\caption{(Color online)
Graphical representations of surviving strategies in $M_2$.
(a) The full transition graph for I-TFT.
Only the black vertices are recurrent states, while others are transient
(see also Fig.~\ref{fig:tran}).
The dashed arcs indicate the paths activated when an error
occurs between I-TFT's. Since two strategies acting as I-TFT are found, we
describe the duality in the graphs by the dotted arcs connected to two
$(CD,CD)$ (at the top-left and the bottom).
(b) Parts of transition graphs characterizing EC and
ET. While an EC-typed strategy tries to recover mutual
cooperation $(CC,CC)$ from an erroneous state $(CC,CD)$ by the
dashed lines, an ET-typed strategy repeatedly defects it
by the dotted line.}
\label{fig:rep}
\end{figure*}

During $1.4\times10^5$ steps to reach the goal, we record the
number of remaining strategies $N$ and their expected payoffs $U$ ranged
over $[U_{\rm min}, U_{\rm max}]$. For comprehension, this range is
divided by $N$ at each step in Fig.~\ref{fig:pc}.
Both of $U_{\rm min}/N$ and $U_{\rm max}/N$ eventually shrinks to a single
point at $3.0$, indicating that all of the strategies obtain $R=3$ from
mutual cooperation.
From the concave shape of $U_{\rm max}/N$, we see two eras:
Roughly before three-quarters of the whole period, $U_{\rm max}/N$
decreases by removing the least fit, as the top-ranked strategies exploit
naive cooperators.
After the prey is consumed out, however, they become the next victims.
Removing defectors now enhances the degree of cooperation and $U_{\rm
max}/N$ rises up as well. There are observed two great extinctions in
that $2^{12}=4096$ strategies disappear simultaneously at the $1424$th and
$124~910$th steps, respectively. These are symbolic of two eras, because AC
is taken off at the first extinction and AD is at the second.

After completing the filtering procedure,
we find an equilibrium, where $12~944$ surviving strategies constitute
a set $\Omega$. The number is still large but only about 5\% of $|M_2|$.
There are two properties in this set:
(i)~All strategies in $\Omega$  are nice in the sense that they never defect
first.
(ii)~After defected at the last two steps, about $94\%$ of the surviving
strategies choose to retaliate.
It is also remarkable that GT and TFT are included in $\Omega$ but Pavlov is
dropped out.

\begin{figure}
\includegraphics[width=.48\textwidth]{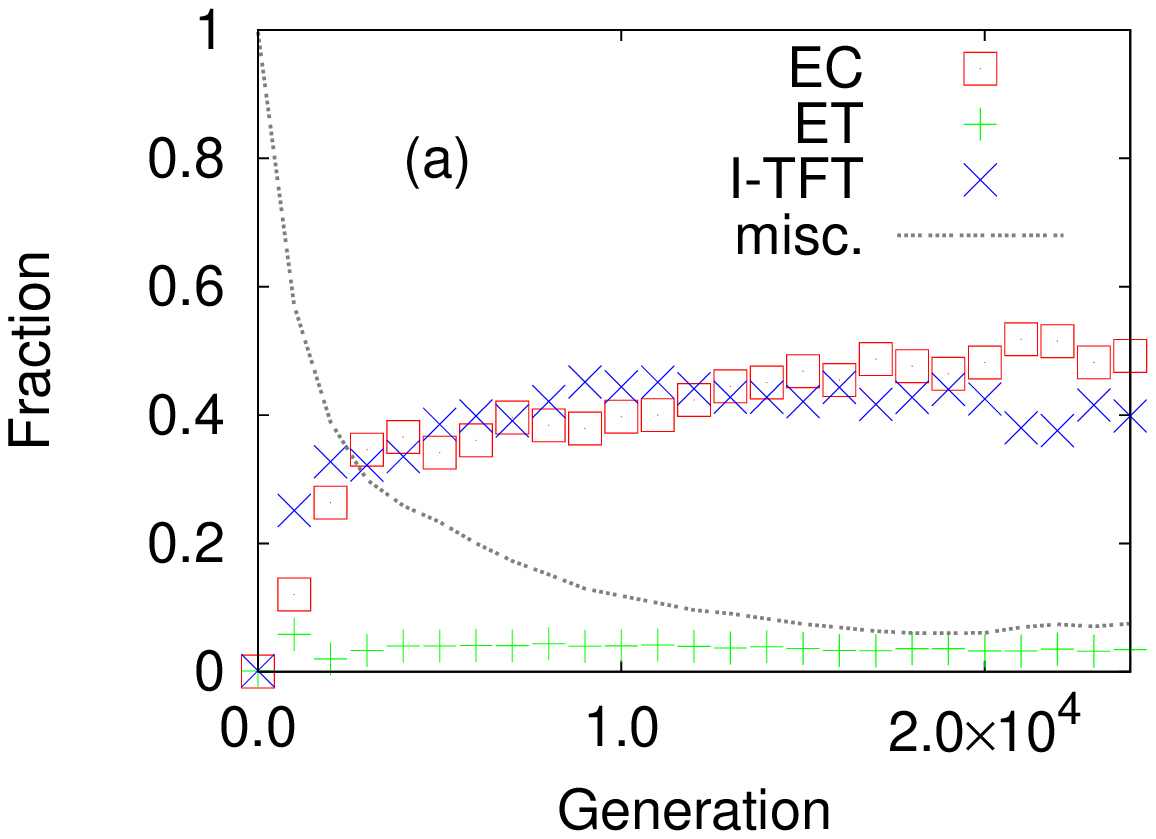}
\includegraphics[width=.48\textwidth]{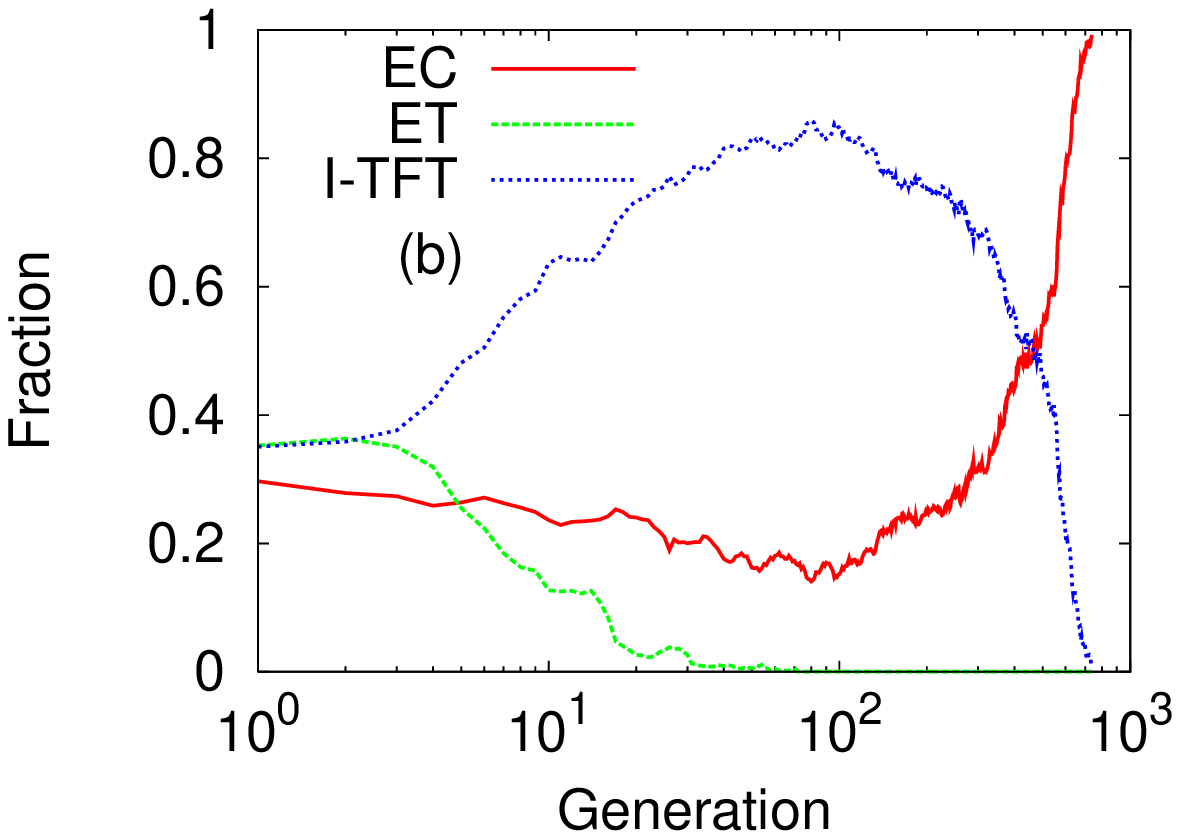}
\includegraphics[width=.48\textwidth]{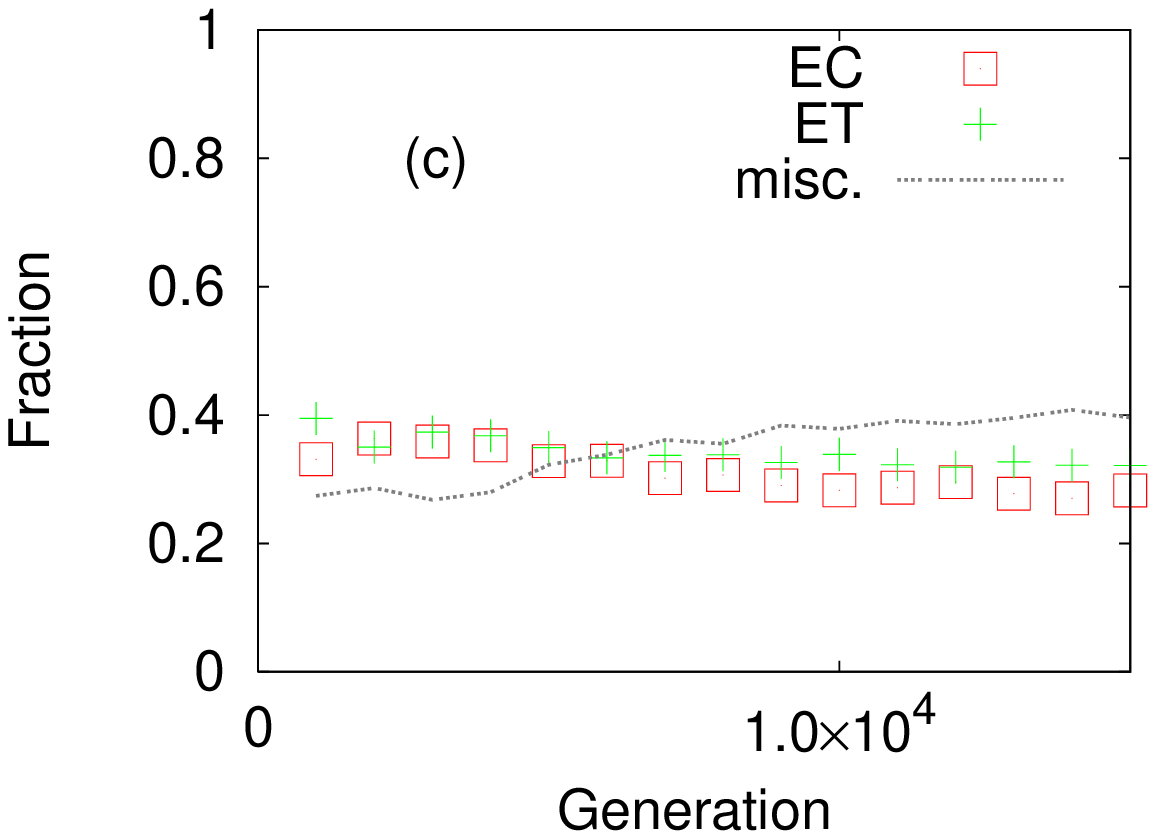}
\caption{(Color online)
SPDG for $M_2$.
(a) All of the $12~944$ strategies in $\Omega$
are initially distributed on $500\times500$
lattices. They are classified as EC, ET, I-TFT and other
miscellaneous ones, and the plotted values are averaged over $50$
realizations.
(b) Three representative strategies belonging to EC,
ET, and I-TFT, respectively, are distributed on a $32 \times 32$
lattice.
(c) Averaged results over 10 realizations on $500\times500$ lattices,
after removing two I-TFT strategies and their six similar variants
(see text) from $\Omega$.
There are given $p=0.02$ and $e=0.01$ in common.
}
\label{fig:m2}
\end{figure}

\subsection{Spatial Prisoner's Dilemma Game for $M_2$}
We next perform SPDG with $e>0$ for $\Omega$ on a two-dimensional
$500\times500$ lattice in the same manner as we did for $M_1$.
Note that the lattice size is almost 20 times greater than the
number of strategies, which turns out to be enough to find
recognizably common patterns.
After $2.4 \times 10^4$ generations, most strategies in $\Omega$ also
disappear and the number of survivors is usually less than 10 in each
realization (Table~\ref{table:str2}).

First, we observe two strategies with only eight recurrent states per each.
They are named as intelligent-TFT (I-TFT) in common, because
TFT is embedded as an attractor in their recurrent states
and the transient states are activated only when an error
occurs~[Fig.~\ref{fig:rep}(a)].
Again, the state $(X_t X_{t+1},Y_t Y_{t+1})$ represents that 
$X_t$ and $X_{t+1}$ ($Y_t$ and $Y_{t+1}$) are the player's (the opponent's) 
moves at two subsequent times ($X,Y = C$ or $D$), which is connected to
$(X_{t+1} X_{t+2},Y_{t+1} Y_{t+2})$ by a directed arc.
Without errors, they are ordinary TFT and never sucked repeatedly by any
other strategies. With errors, on the other hand, they return back to mutual
cooperation without the chain retribution between themselves,
overcoming the weakness of the classical TFT. Furthermore, this
error tolerance is secured from repeated abuse by being transient.
We therefore conclude that the only way to defeat I-TFT is more
efficient cooperation than I-TFT's.
As long as the error occurs rarely enough not to disturb its recovery path,
I-TFT will clear the defecting strategies out and eventually make way for
better cooperators.

Such efficient cooperators are characterized by the way of dealing with an
error between themselves, depicted in Fig.~\ref{fig:rep}(b) with the
dashed lines.
We denote those strategies with such an error recovery path as efficient
cooperator (EC).
An EC-typed strategy outperforms I-TFT because it costs less by one point in
recovering an error. This one point may look small but has a significant
meaning after thousands of generations.
Yet an EC strategy can be invaded by even such a trivial
strategy in $M_1$ as $\alpha|DDCC$ which simply alternate between $C$ and
$D$, regardless of $\alpha$. That is,
inserted among the strategies of $M_1$, an EC-typed strategy does not
overwhelm $M_1$ and sometimes becomes exterminated. Meanwhile, I-TFT under
the same condition works so successfully that it wins the whole area by
defeating all of the $M_1$ strategies, including Pavlov, in every
realization so long as $p$ is small enough.

Last, some cooperating strategies are triggered to deceive EC by a single
error: At the last step of EC's error recovery phase,
they defect again, instead of getting back to $(CC,CC)$ as desired,
and complete the exploiting loop [see the dotted line in
Fig.~\ref{fig:rep}(b)]. Even if they are trigger strategies specialized to
defeat EC, from which we simply call them efficient trigger (ET), I-TFT
suppresses them and helps EC to rise [Figs.~\ref{fig:m2}(a) and
\ref{fig:m2}(b)].

Those two I-TFT strategies are distinguished by how they respond to
the state $(CD,CD)$ (Table~\ref{table:str2}). Let us denote the I-TFT
strategy responding with $C$ as I-TFT$_C$ and that with $D$ as I-TFT$_D$.
Comparing a population of I-TFT$_C$ with that of I-TFT$_D$, the former
is slightly better off, as the latter has a probability of $O(e^2)$ that
both players make errors at the same time, leading to $(CD,CD) \rightarrow
(DD,DD)$ [Fig.~\ref{fig:rep}(a)].

If we repeat this SPDG procedure after removing I-TFT from $\Omega$,
some variants of I-TFT play the role of protecting EC. They have only one
or two different bits from either of I-TFT strategies, but their
recurrent states do not constitute TFT. Further removing such variants,
we see that EC strategies are helplessly threatened by the parasitic ET
[Fig.~\ref{fig:m2}(c)]. Since ET strategies cannot do well with errors, the
level of cooperation remains low. This comparison clearly shows the crucial
role of I-TFT.

Let us recall Pavlov in comparison with EC:
While GT and TFT ignored the presence of an error within the same
species, not to be sucked by anyone, Pavlov invented a
recovery path $(C,D) \rightarrow (D,D) \rightarrow (C,C)$ and could be
the final winner in $M_1$. Nevertheless, it is at the very
point that GT overruns Pavlov. It is therefore not surprising that Pavlov
fails to enter $\Omega$, because so many strategies of $M_2$ are willing to
exploit its shortsighted tolerance. Even though EC devises a more
sophisticated recovery path than Pavlov's, it is still far from safe.
The point is that all of their states are recurrent: Even if they use every
given memory capacity to determine the next move, once the patterns are
recognized, the opponent can get back to the defecting state as many times
as it wants.
However, EC strategies are successful in the long run, because they try
to cooperate better at some expense of security risk. The success of EC
crucially depends on the existence of such balancing strategies as I-TFT,
and is thus path-dependent.

\section{Summary}
\label{sec:summary}
In summary, we presented a thorough examination on strategies under
restrictions of the time horizon in the iterated PD game.
As the time horizon is enlarged, a variety of trajectories to equilibrium
become possible, but there are still common dynamical patterns.
That is, the system reaches efficient cooperation through intermediate
prevalence of TFT-like strategies, which solve the dilemma between security
and tolerance by using transient states.
As I-TFT spends most time as the classical TFT which refers only
to the opponent's last move, it becomes even more likely to win
if memory is costly~\cite{Sethi}. 
This gives a clue for understanding how the
memory could be effectively saved in social interactions and differentiated
into other functions.

The detailed features of our observation in this paper may be partially
owing to our specific choice of elementary payoff values. However, we
believe that the successful strategies such as I-TFT and dynamical
patterns between them have good reasons to be remarkable in a more general
context of the evolutionary PD game.

\acknowledgments
We are thankful to Jung-Kyoo Choi for discussions.
This work was supported by the Korea Science and Engineering Foundation
Grant No. R01-2007-000-20084-0.

\bibliographystyle{revtex}

\begin{thebibliography}{10}
\providecommand*{\bibinfo}[2]{#2}
\providecommand*{\eprint}[1]{#1}
\providecommand*{\url}[1]{#1}
\bibitem{Newman}
\bibinfo{author}{M.~E.~J. Newman} and \bibinfo{author}{G.~T. Barkema},
  \bibinfo{title}{\emph{Monte Carlo Methods in Statistical Physics}}
  (\bibinfo{publisher}{Oxford University Press}, New York,
  \bibinfo{year}{1999}).
\bibitem{Nash}
\bibinfo{author}{J.~Nash}, \bibinfo{journal}{Proc. Natl. Acad. Sci. U.S.A.}
  \bibinfo{volume}{\textbf{36}}, \bibinfo{pages}{48} (\bibinfo{date}{1950}).
\bibitem{Axelrod}
\bibinfo{author}{R.~Axelrod}, \bibinfo{title}{\emph{The Evolution of
  Cooperation}} (\bibinfo{publisher}{Basic Books}, New York,
  \bibinfo{year}{1984}).
\bibitem{Roca}
\bibinfo{author}{C.~P. Roca}, \bibinfo{author}{J.~A. Cuesta}, and
  \bibinfo{author}{A.~S\'anchez}, \bibinfo{journal}{Phys. Rev. Lett.}
  \bibinfo{volume}{\textbf{97}}, \bibinfo{pages}{158701}
  (\bibinfo{date}{2006}).
\bibitem{Friedman}
\bibinfo{author}{J.~W. Friedman}, \bibinfo{journal}{Rev. Econ. Stud.}
  \bibinfo{volume}{\textbf{38}}, \bibinfo{pages}{1} (\bibinfo{date}{1971}).
\bibitem{Kraines}
\bibinfo{author}{D.~Kraines} and \bibinfo{author}{V.~Kraines},
  \bibinfo{journal}{Theory Decis.} \bibinfo{volume}{\textbf{26}},
  \bibinfo{pages}{47} (\bibinfo{date}{1989}).
\bibitem{Leimar}
\bibinfo{author}{O.~Leimar}, \bibinfo{journal}{J. Theor. Biol.}
  \bibinfo{volume}{\textbf{184}}, \bibinfo{pages}{471} (\bibinfo{date}{1997}).
\bibitem{Axelrod1987}
\bibinfo{author}{R.~Axelrod}, in \emph{Genetic Algorithms and Simulated
	Annealing} edited by L. Davis (\bibinfo{publisher}{Morgan
  Kaufmann}, Los Altos, California, \bibinfo{year}{1987}).
\bibitem{Miller}
\bibinfo{author}{J.~H. Miller}, \bibinfo{journal}{J. Econ. Behav. Organ.}
  \bibinfo{volume}{\textbf{29}}, \bibinfo{pages}{87} (\bibinfo{date}{1996}).
\bibitem{Lindgren1994}
\bibinfo{author}{K.~Lindgren} and \bibinfo{author}{M.~G. Nordahl},
  \bibinfo{journal}{Physica D} \bibinfo{volume}{\textbf{75}},
  \bibinfo{pages}{292} (\bibinfo{date}{1994}).
\bibitem{Nowak1990}
\bibinfo{author}{M.~Nowak}, \bibinfo{journal}{Theor. Popul. Biol.}
  \bibinfo{volume}{\textbf{38}}, \bibinfo{pages}{93} (\bibinfo{date}{1990}).
\bibitem{Nowak1992a}
\bibinfo{author}{M.~Nowak} and \bibinfo{author}{K.~Sigmund},
  \bibinfo{journal}{Nature} \bibinfo{volume}{\textbf{355}},
  \bibinfo{pages}{250} (\bibinfo{date}{1992}).
\bibitem{Nowak1993}
\bibinfo{author}{M.~Nowak} and \bibinfo{author}{K.~Sigmund},
  \bibinfo{journal}{Nature} \bibinfo{volume}{\textbf{364}}, \bibinfo{pages}{56}
  (\bibinfo{date}{1993}).
\bibitem{Hauert}
\bibinfo{author}{C.~Hauert} and \bibinfo{author}{H.~G. Schuster},
  \bibinfo{journal}{Proc. R. Soc. London, Ser. B} \bibinfo{volume}{\textbf{264}},
  \bibinfo{pages}{513} (\bibinfo{date}{1997});
  \bibinfo{journal}{J. Theor. Biol.} \bibinfo{volume}{\textbf{192}},
  \bibinfo{pages}{155} (\bibinfo{date}{1998}).
\bibitem{killing}
\bibinfo{author}{T.~Killingback} and \bibinfo{author}{M.~Doebeli},
  \bibinfo{journal}{Am. Nat.} \bibinfo{volume}{\textbf{160}},
  \bibinfo{pages}{421} (\bibinfo{date}{2002}).
\bibitem{spatial}
\bibinfo{author}{M.~Nowak} and \bibinfo{author}{R.~M. May},
  \bibinfo{journal}{Nature} \bibinfo{volume}{\textbf{359}},
  \bibinfo{pages}{826} (\bibinfo{date}{1992});
\bibinfo{author}{G.~Szab\'o} and \bibinfo{author}{C.~T{\"o}ke},
  \bibinfo{journal}{Phys. Rev. E} \bibinfo{volume}{\textbf{58}},
  \bibinfo{pages}{69} (\bibinfo{date}{1998});
\bibinfo{author}{M.~A. Nowak}, \bibinfo{journal}{Science}
  \bibinfo{volume}{\textbf{314}}, \bibinfo{pages}{1560} (\bibinfo{date}{2006}).
\bibitem{network}
\bibinfo{author}{B.~J. Kim}, \bibinfo{author}{A.~Trusina},
  \bibinfo{author}{P.~Holme}, \bibinfo{author}{P.~Minnhagen},
  \bibinfo{author}{J.~S. Chung}, and \bibinfo{author}{M.~Y. Choi},
  \bibinfo{journal}{Phys. Rev. E} \bibinfo{volume}{\textbf{66}},
  \bibinfo{pages}{021907} (\bibinfo{date}{2002});
\bibinfo{author}{P.~Holme}, \bibinfo{author}{A.~Trusina},
  \bibinfo{author}{B.~J. Kim}, and \bibinfo{author}{P.~Minnhagen},
  \bibinfo{journal}{{\it ibid.}} \bibinfo{volume}{\textbf{68}},
  \bibinfo{pages}{030901(R)} (\bibinfo{date}{2003});
\bibinfo{author}{H.~Ohtsuki}, \bibinfo{author}{C.~Hauert},
  \bibinfo{author}{E.~Lieberman}, and \bibinfo{author}{M.~A. Nowak},
  \bibinfo{journal}{Nature} \bibinfo{volume}{\textbf{441}},
  \bibinfo{pages}{502} (\bibinfo{date}{2006});
\bibinfo{author}{J.~G\'omez-Garde{\"n}es}, \bibinfo{author}{M.~Campillo},
  \bibinfo{author}{L.~M. Flor\'ia}, and \bibinfo{author}{Y.~Moreno},
  \bibinfo{journal}{Phys. Rev. Lett.} \bibinfo{volume}{\textbf{98}},
  \bibinfo{pages}{108103} (\bibinfo{date}{2007}).
\bibitem{szabo}
\bibinfo{author}{G.~Szab\'o} and \bibinfo{author}{G. F\'ath},
  \bibinfo{journal}{Phys. Rep.} \bibinfo{volume}{\textbf{446}},
  \bibinfo{pages}{97} (\bibinfo{date}{2007}).
\bibitem{sheng}
\bibinfo{author}{Z.-X.~Wu}, \bibinfo{author}{X.-J.~Xu},
	\bibinfo{author}{Z.-G. Huang},
	\bibinfo{author}{S.-J.~Wang}, and \bibinfo{author}{Y.-H.~Wang},
  \bibinfo{journal}{Phys. Rev. E} \bibinfo{volume}{\textbf{74}},
  \bibinfo{pages}{021107} (\bibinfo{date}{2006});
\bibinfo{author}{Z.-H.~Sheng}, \bibinfo{author}{Y.-Z.~Hou},
	\bibinfo{author}{X.-L.~Wang}, and \bibinfo{author}{J.-G.~Du},
  \bibinfo{journal}{J. Phys.: Conf. Ser.} \bibinfo{volume}{\textbf{96}},
  \bibinfo{pages}{012107} (\bibinfo{date}{2008}).
\bibitem{Ashlock}
\bibinfo{author}{D.~Ashlock}, \bibinfo{author}{M.~D. Smucker},
  \bibinfo{author}{E.~A. Stanley}, and \bibinfo{author}{L.~Tesfatsion},
  \bibinfo{journal}{BioSystems} \bibinfo{volume}{\textbf{37}},
  \bibinfo{pages}{99} (\bibinfo{date}{1996}).
\bibitem{Sethi}
\bibinfo{author}{R.~Sethi} and \bibinfo{author}{E.~Somanathan},
  \bibinfo{journal}{J. Econ. Behav. Organ.} \bibinfo{volume}{\textbf{50}},
  \bibinfo{pages}{1} (\bibinfo{date}{2003}).
\bibitem{Weibull}
\bibinfo{author}{J.~M. Weibull}, \bibinfo{title}{\emph{Evolutionary Game
  Theory}} (\bibinfo{publisher}{MIT Press}, Cambridge,
  \bibinfo{year}{1995}).
\bibitem{Kandori}
\bibinfo{author}{M.~Kandori}, \bibinfo{author}{G.~J. Mailath}, and
  \bibinfo{author}{R.~Rob}, \bibinfo{journal}{Econometrica}
  \bibinfo{volume}{\textbf{61}}, \bibinfo{pages}{29} (\bibinfo{date}{1993}).
\bibitem{Lindgren1992}
\bibinfo{author}{K.~Lindgren}, in
  \emph{Artifical Life II}, edited by
  \bibinfo{editors}{C.~G. Langton, C.~Taylor, J.~D. Farmer, and S.~Rasmussen}
  (\bibinfo{publisher}{Addison-Wesley}, Redwood City, CA,
  \bibinfo{year}{1992}).
\bibitem{Tanimoto}
\bibinfo{author}{J.~Tanimoto} and \bibinfo{author}{H.~Sagara},
  \bibinfo{journal}{BioSystems} \bibinfo{volume}{\textbf{90}},
  \bibinfo{pages}{728} (\bibinfo{date}{2007}).
\bibitem{Borg}
\bibinfo{author}{T.~B{\" o}rgers}, \bibinfo{journal}{Econometrica}
  \bibinfo{volume}{\textbf{61}}, \bibinfo{pages}{423} (\bibinfo{date}{1993}).
\bibitem{cycle}
\bibinfo{author}{M.~Nowak} and \bibinfo{author}{K.~Sigmund},
  \bibinfo{journal}{J. Theor. Biol.} \bibinfo{volume}{\textbf{137}},
  \bibinfo{pages}{21} (\bibinfo{date}{1989});
\bibinfo{author}{J.~Hofbauer} and \bibinfo{author}{K.~Sigmund},
\bibinfo{title}{\emph{Evolutionary Games and Population Dynamics}}
(\bibinfo{publisher}{Cambridge University Press}, Cambridge,
  \bibinfo{year}{1998}).

\end{thebibliography}

\end{document}